%% file: MRRMayzodi3.tex
\documentstyle[epsfig]{mn2e}
\begin{document}
\input BoxedEPS.tex
\input macro.tex
\SetEPSFDirectory{/scratch/sbgs/figures/hst/}
\SetRokickiEPSFSpecial
\HideDisplacementBoxes

\title[Infrared emission from the zodiacal dust cloud]{An improved model for the infrared emission from the zodiacal dust 
cloud: cometary, asteroidal and interstellar dust}
\author[Rowan-Robinson M. and May B.]{Michael Rowan-Robinson and Brian May\\
Astrophysics Group, Blackett Laboratory, Imperial College of Science 
Technology and Medicine, Prince Consort Road,\\ 
London SW7 2AZ,\\
}
\maketitle
\begin{abstract}
We model the infrared emission from zodiacal dust detected by the IRAS and COBE missions,
with the aim of estimating the relative contributions of asteroidal, cometary and interstellar dust
to the zodiacal cloud.  Our most important result is the detection of an isotropic component of foreground
radiation due to interstellar dust.

The dust in the inner solar system is known to have a fan-like distribution.  If this is assumed to extend to
the orbit of Mars, we find that cometary, asteroidal and interstellar dust account for 70$\%$, 22$\%$
and 7.5$\%$ of the dust in the fan.  We find a worse fit if the fan is assumed to extend
 to the orbit of Jupiter.    Our model is broadly consistent with 
 the analysis by Divine (1993) of interplanetary 
dust detected by Ulysses and other spacecraft.  Our estimate of the mass-density of interstellar dust in the inner solar system is
consistent with estimates from Ulysses at 1.5 au, but is an order of magnitude higher than Ulysses estimates
at r $>$ 4 au.  Only 1$\%$ of the zodiacal dust arriving at the earth would be interstellar, in our model.

Our models can be further tested by ground-based kinematical studies of the zodiacal cloud, which need to
extend over a period of years to monitor solar cycle variations in interstellar dust, by dynamical simulations, 
and by in situ measurements from spacecraft.

\end{abstract}
\begin{keywords}
Planetary systems - zodiacal dust, interplanetary medium.
\end{keywords}


\section{Introduction}
The zodiacal dust is an important constituent of the Sun's debris disk, and contains clues to the recent history 
of that disk.  In this paper we attempt to estimate the relative contributions of asteroidal, cometary and interstellar
dust to the infrared emission from zodiacal dust detected by IRAS and COBE.  This is the first attempt
to model the data from both missions simultaneously.  

\subsection{Pre-IRAS}
Early work on modelling scattered optical zodiacal light was extensively reviewed by Giese et al
(1985, 1986) and Leinert (1985).  The consensus from these studies was that the number density
of dust grains in the inner solar system followed a distribution\\

\medskip
	$n(r) = n_0 r^{-\gamma} f(\beta_0)$	(1)\\

where r is the distance from the Sun in A.U. and $\beta_0$ is the elevation from the symmetry plane
of the dust, with $\gamma \sim$ 1.3.

 A popular form for $f(\beta_0)$ was the 'fan' distribution\\ 

\medskip
	$f(\beta_0) = exp - P |sin \beta_0|$.	(2)\\

\subsection{Kinematical studies}
Early work on the kinematics of the zodiacal dust was begun by the Imperial group in the late 1960s
 (Reay and Ring 1968, Hicks, May and Reay 1972, 1974, East and Reay 1984).
Hicks et al (1974) raised the possibility that some of the zodiacal dust in the vicinity of earth may be of
interstellar origin, an idea that was developed more fully by May (2007).  May's estimate was that
the kinematic asymmetry seen by Hicks et al (1974) between Sept-Oct 1971 and April 1972 could be explained if
$\sim10\%$ of the dust were part of a linear flow, due for example to interstellar dust, with the rest being 
due to dust in circular orbits, arising from cometary and asteroidal sources.

More recent kinematic measurements of zodiacal dust have been made by Reynolds et al (2004), using the WHAM
(Wisconson H-alpha Machine) spectrometer, and these observations have been modelled by Madsen et al (2006) 
using models of cometary and
asteroidal dust published by Ipatov et al (2006).  However May (2007) still found that there was a discrepancy
between the observations of Reynolds et al (2004) and those of East and Ray (1984), which might still leave room
for a seasonal variation in zodiacal dust kinematics due to the presence of interstellar dust.

\subsection{IRAS}
Our understanding of the zodiacal dust was transformed by the IRAS mission in 1983, which gave the first all-sky 
maps of the zodiacal dust emission and discovered both the  ÔIRAS Zodiacal BandsÕ, attributed to collisions between members
of asteroid families (Low et al 1984, Dermott et al 1984, Sykes and Greenberg 1986, Sykes 1990, Grogan et al 2001), of which at 
least 5 are now known, contributing about 10 per cent of the total zodiacal emission in the infrared, and, 
on a small scale, cometary dust trails (Sykes et al 1986).  These demonstrated that both 
collisions between asteroids and comet debris must, at some level, contribute to the zodiacal dust cloud.   
IRAS also revealed a local ring of dust near the Earth (Dermott et al 1994, 
Leisawitz et al 1994, Reach et al 1995).

The first full models of infrared emission from the zodiacal dust cloud were given by Rowan-Robinson et al 
1990, 1991, Jones and Rowan-Robinson 1993.  
The novel features of these models were:

(i) use of a modified fan distribution\\

\medskip
	$f(\beta_0) = (cos \beta_0)^Q  exp - P sin |\beta_0|^{\xi}$.	(3)\\

This incorporates the $(cos \beta_0)^Q$ term introduced by Murdock and Price (1985) and
the smoothing at the symmetry plane introduced by Deul and Wolstencroft (1988), where\\

\medskip
$\xi = 2 - |z/z_0|$, for $|z| < z_0$, = 1 otherwise,\\ 
and $z_0$ = 0.065 au.\\
Dust grains were assumed to be grey ($Q_{\nu}$ = 1) at 12-100 $\mu$m, so their typical radii needed to be
$> 10 \mu$m.

(ii) detailed models of the two strongest pairs of narrow asteroid band features at $|\beta$ = 1.4 and 10.2$^o$, 

(iii) identification of an additional component of broad asteroidal bands centred at $|\beta| \sim 10^o$.
Rowan-Robinson et al (1991) suggested that the cooler spectral energy distribution of the broad
bands pointed to a distant origin for these bands, perhaps at $\sim$ 50au, but a parallax test
performed by Jones and Rowan-Robinson (1993) using the 3rd IRAS HCON, showed they are
consistent with being, like the narrow bands, at r $\sim$1.5-3 au.
Jones and Rowan-Robinson's (1993) fan model extended to r = 1.5 au and they found that asteroidal dust 
could account for $\sim 25 \%$ of the dust involved in the fan component.  So a major deficiency of that
model is to account for the remaining $\sim 75 \%$ of the dust in the fan.

Liou et al (1995) used a dynamical analysis fitted to the IRAS data to conclude that 74$\%$ of the zodiacal cloud
was of cometary origin, while 26$\%$  was asteroidal.  Durda and Dermott (1997) concluded that as much as
34$\%$ could be asteroidal.

Nesvorny et al (2010) have modelled the dynamics of cometary dust and compared results with IRAS data,
concluding that over 90$\%$ of zodiacal dust is cometary.
However their use of smoothed data means that the narrow asteroidal bands are smoothed out and it is likely that they
have underestimated the contribution of asteroidal dust.

\subsection{Ulysses}
The next major contribution to understanding of the nature and origin of zodiacal dust was in situ 
measurements of zodiacal dust by Ulysses (1990-2009) and other spacecraft.
Divine (1993) analysed Ulysses and other data to identify five different components
of interstellar dust.  The three dominant components are the core (fan) component, extending to
$\sim$ 2 au, an asteroidal component dominating from 1-3 au and a relatively isotropic halo component
which becomes significant at $\sim$ 3 au and dominates at large distances (see also the review of solar
 system dust by Grun et al 1993).

One of the major discoveries of Ulysses was the strong role of interstellar dust in the solar system,
especially at greater distances from the Sun (see the review by Mann 2010).  Interstellar dust has also been 
detected in the solar system by the Galileo, Cassini and Stardust spacecraft (Altobelli et al 2003, Grun et al 2005, 
Mann 2010, Kruger et al 2010, Sterken et al 2012a).   The density of
interstellar dust in the solar system varies with time, with a strong dependence on the solar cycle.
Because the dust particles become charged, their motion is strongly influenced by the plasma
flow, and the earth's bow shock excludes much of the dust, especially the smaller particles.
At times of solar maximum, magnetic field reversals mean that interstellar dust is allowed
into the inner solar system.  Its relatively high velocity relative to the Sun, typically $\sim 26 km s^{-1}$,
means that this dust can traverse the $\sim$ 200 au from the magnetopause in $\sim$ 50 years.
By contrast dust in approximately circular orbit around the sun, and spiralling inwards under
the influence of the Poynting-Robertson effect, takes $\sim$10,000 years to travel from the asteroid belt to the Earth
(Gustafsen et al 1987).

Interstellar dust within a cylindrical column with the Hoyle-Lyttleton radius $2GM_{\odot}/v_{\odot}^2 \sim$ 4 au 
(Hoyle and Lyttleton 1939) would be gravitationally
trapped by the Sun and can in principal be an important supply source for the zodiacal dust cloud. Since the time-scale 
for this gravitational trapping is of order  $\sim$ 2 years, the quantity of interstellar dust in the inner solar system
may vary on a time-scale of years, and would certainly be expected to vary over the solar cycle.
Interstellar dust with impact parameter appreciably greater than 4 au would pass through the inner solar system rapidly 
but may still contribute to the infrared emission detected by IRAS and DIRBE.  

The planets have a negligible effect on the motion of the interstellar dust.  By contrast the very slow passage
of dust grains spiralling inwards under the Poynting-Robertson effect ensures that each planet can
significantly modify the orbits of dust grains and the vertical distribution of dust density.

Grogan et al (1996) simulated the flow of interstellar dust into the inner solar system, including the effects of gravity, 
radiation pressure, and magnetic forces on the dust grains.  They found that the distribution of this dust tends to
be approximately isotropic over most of the sky, with a column downstream of the flow direction in which larger grains ($\gg 0.3 \mu$m)
are focussed by the Hoyle-Lyttleton accretion, and from which smaller grains ($< 0.3 \mu$m) are excluded by the 
combined effects of radiation pressure and magnetic forces.  They concluded that the contribution of local interstellar
grains to the infrared foreground would be small ($< 0.1 MJy sr^{-1}$ at 12 $\mu$m), but this was based on rather
a low estimate of the density of interstellar dust at 5 a.u. (see section 6 below).  A more detailed simulation of the flow
of interstellar dust through the solar system, under the action of gravity, radiation pressure and Lorentz forces, through the 
different phases of the solar cycle, has been given by Sterken et al (2012b).  They find strong variation with time of  the
density of interstellar dust in the inner solar system.  They also find some systematic deviation from isotropy with time 
during the solar cycle, but this is unlikely to be detectable in  the type of modelling we are carrying out here.

\subsection{COBE}
The DIRBE instrument on COBE mapped the zodiacal emission in 1989-90 with a different scan strategy to IRAS.  Whereas
IRAS scanned from ecliptic pole to pole, the COBE spacecraft executed an additional circular motion which resulted in sampling
the zodiacal emission at a wide range of solar elongations every day.  A further advantage was that DIRBE, unlike IRAS,
had an absolute calibration.  The IRAS zodiacal measurements had a somewhat better resolution than
DIRBE, and coupled with the fact that scans were at approximately constant solar elongation, this meant that
IRAS gives better resolution of the fine structure in the zodiacal bands.
 
Kelsall et al (1998) modelled the zodiacal emission detected by DIRBE using a slightly different modified exponential 
fan to JRR with\\

\medskip
	$f(\beta_0) = exp -P (sin |\beta_0| - z_0/2)^{\gamma_1}$, for $ sin |\beta_0| <z_0$,\\
\medskip	
			$= exp -P ( sin^2  |\beta_0| /2 z_0)^{\gamma_1}$, for $ sin |\beta_0| > z_0$,\\
\medskip			
			where $z_0$ = 0.189, $\gamma_1$ = 0.942.\\
\medskip	
	
No attempt was made to account for cometary 
or interstellar dust, but the fan is assumed to extend to 5.2 AU.  Their main innovation is inclusion of the ring of dust at
r $\sim$1au and the blob trailing behind earth, which had been first detected by Dermott et al (1994) in the IRAS data,
and confirmed by Reach et al (1995).  They justified the neglect of interstellar dust based on the work of
Grogan et al (1996).

In this paper we revisit models for the combined IRAS and DIRBE observations of zodiacal dust, with a view to trying to
determine the relative contributions of asteroidal, cometary and interstellar dust.  The IRAS
Zodiacal History File remains a key data set if full sampling of the asteroidal component is
required.   The DIRBE data, with its absolute calibration, is essential if an isotropic component like
interstellar dust is to be detected. The components in our model are physically motivated, occupy distinct regions 
of the solar system, and have different radial dependences.  While dynamical simulations are essential for 
understanding the evolution of zodiacal dust, our approach provides a valid independent perspective.

\section{Ingredients of the model}
We aim to be strongly guided by the evidence from direct detection of interplanetary dust.
We can expect major transformation of the vertical dust profile as each planetary orbit is crossed and so for
our purposes we must allow for significant changes at the orbit of Mars, r = 1.53 au, and at the orbit of Jupiter,
r = 5.2 au.

The main ingredients of the Jones and Rowan-Robinson (1993, JRR) model, the modified fan and asteroid bands, 
correspond well to two of the main components detected in spacecraft data.  Their assumption that the fan 
terminates at r = 1.5 au is consistent with the Divine (1993) analysis, though we will explore here the possibility
that it extends further.  The radial dependence for the fan assumed by Jones and Rowan-Robinson was $r^{-1.0}$, 
guided by the expected behaviour of grains subject to the Poynting-Robertson effect, but the evidence for a
radial dependence of $r^{-1.3}$ within 1.5 au now seems very strong and we use that dependence here.
We initially use the JRR parameters for the asteroid bands, but will consider the possibility that the amplitudes need
modification, to take account of the other ingredients used here.  We will also look briefly at the possibility
that the broad bands extend to a greater distance than 3.1 au.

The first ingredient not incorporated by JRR, but included by Kelsall et al (1998), is the solar ring and trailing
blob and we have used the formalism of Kelsall et al (1998) for these, but allowing the amplitudes to be a
free parameter.

Secondly we want to include a component corresponding to cometary dust.  The dominant contribution is
expected to be from Jupiter-family comets, which tend to have inclinations $<$ 30$^o$, have aphelia ranging 
from 5 to 30 au, and have their origin
in the Kuiper belt.  For the cometary contribution to the asteroidal dust we therefore expect a flattened 
distribution, but perhaps not so strongly flattened as the inner fan.  We assume an exponential
fan extending from r = 1.5 to 30 au, with exponent PCOM (to be determined), and assuming an $r^{-1}$ radial distribution.  
Nesvorny et al (2010) estimate that the contribution of Oort cloud comets is negligible and we were also not able
to detect such a component.

Finally for interstellar dust we assume an isotropic distribution, with uniform density, extending from r = 0 to 30 au. 
In reality the distribution of interstellar dust will be more complex, especially for larger and smaller grains
(Sterken et al 2012), but isotropy is a reasonable first-order assumption in this attempt to detect interstellar
dust through its infrared emission.

We assume the cometary and interstellar components together contribute the 'halo' population identified by 
Divine (1993).  The crucial distinction between the components in the model lies in their radial distribution,
with asteroidal dust originating between 1.5 and 3.1 au, cometary dust extending from 1.5 au out to large
distances, and interstellar dust extending through the whole solar system.  The fan is assumed to be supplied by
the cometary and asteroidal components and extends to 1.5 au. So the components are quite
distinct in their contribution to the infrared emission.

\section{Fits to IRAS data}

The data we have used is version 3.0 of the IRAS Zodiacal History File, which lists position on the sky, UTCS and fluxes averaged 
over 0.5 deg sq in the four IRAS bands.  From this information the solar longitude can be calculated.  Data heavily
affected by cirrus is excluded, using a standard mask.  The fan is
assumed to have a symmetry plane specified by ($\Omega$, i).  We have not corrected for the relatively small effect
of the Sun's displacement from the symmetry plane of the zodiacal dust cloud.  The density of the dust in the fan is
specified by equations (1) and (3).  The dust grains are assumed to be large and grey, with a temperature
dependence   $T_1 r^{-0.5}$, with $T_1$  = 255 K.  Variation of the temperature at 1 au, $T_1$ , was explored but 
not found to improve the fits.  The fan was assumed to extend out to a radius RMAX, initially taken to be 1.53 au.

The density of dust in the narrow bands were assumed to satisfy equation (1), with $\gamma$ = 1.0 and\\

\medskip
	$f_{nb}( \beta_0) = exp G(|\beta_0 | - \beta_{nb})$, $|\beta_0| < \beta_{nb}$           (4).\\

$\beta_{b}$ was taken to be 1.42$^o$ for the inner bands, corresponding to the Themis asteroid family.  For the
outer bands, JRR assumed $\beta_{b}$ = 10.14$^o$, corresponding to the Eos family, but Grogan et al (2001)
have shown that there is better agreement with the Veritas family, with $\beta_{b}$ = 9.35$^o$, and we confirm
here that this gives a better fit to the IRAS data.  The bands are assumed to extend from r =1.53 to 3.1 au.

For the broad bands identified by Rowan-Robinson et al (1991) and modelled by JRR, we assume they have their
own symmetry plane characterised by ($\Omega_{bb}, i_{bb}$).  Their density is also assumed to be of the
form of equation (1), with $\gamma$ =1.0 and\\

\medskip
	$f_{bb}( \beta_1) = {exp[-(\beta_1 - \beta_{bb})^2/2\sigma_{bb}^2] + exp[-(\beta_1 + \beta_{bb})^2/2\sigma_{bb}^2]}$   (5)\\

where $\beta_{bb}$ is the latitude of the peaks above the symmetry plane and $\sigma_{bb}$ is the peak width.
These bands are assumed to extend from r = 1.53 au to R2BF, where R2BF is normally taken to be 3.1 au, but
the possibility of extension to greater distance is considered below.  The grains in the broad bands were also assumed
to be large and we did not confirm the suggestion of JRR that the fits are improved if the broad band grains are
assumed to be small.

For cometary dust we assume a density dependence of the form equations (1) and (2), with $\gamma$=1.0.
We are assuming that cometary dust has a flattened distribution, but not as strongly flattened as the main
zodiacal fan.  Cometary dust is assumed to extend from r = 1.53 to 30 au.  

\begin{figure*}
\epsfig{file=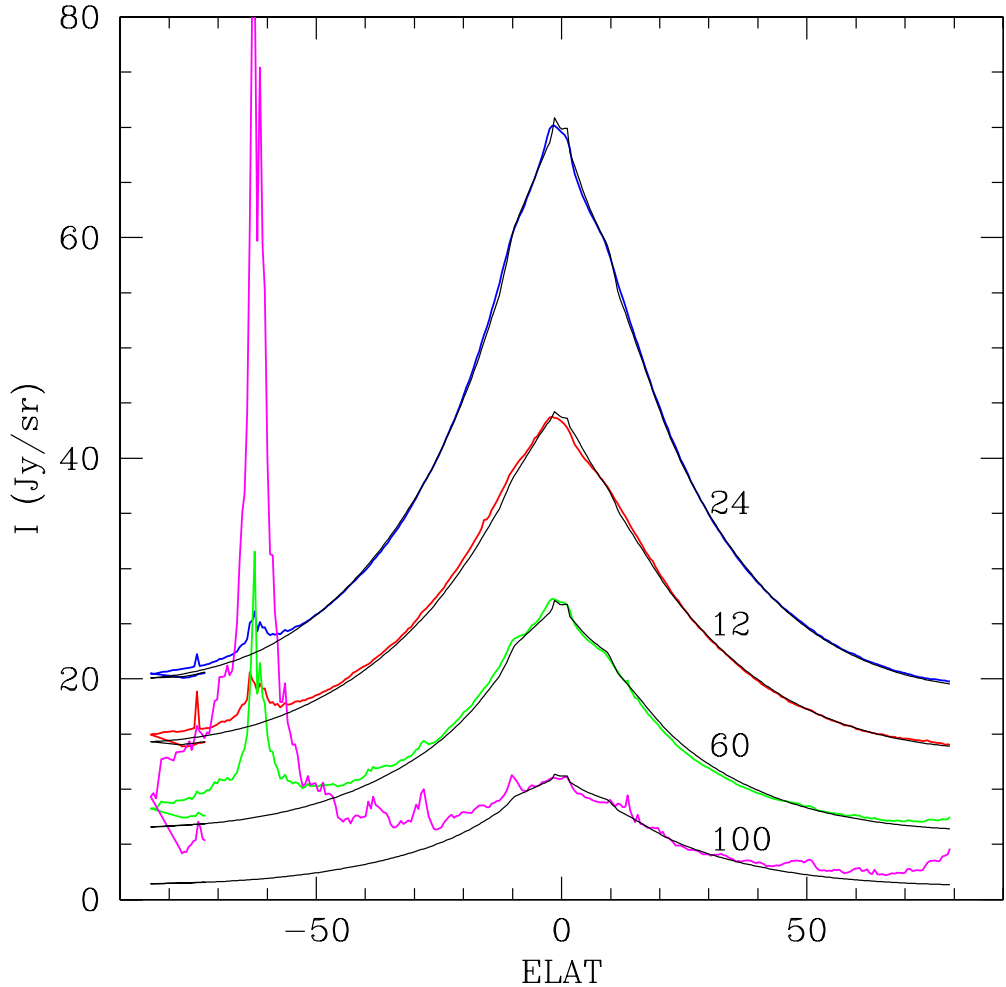,angle=0,width=14cm}
\caption{Comparison of IRAS scans at 12, 25, 60 and 100 $\mu$m with zodiacal dust model A.  Scan is
centred at solar longitude 90.61$^o$.
}
\end{figure*}

For interstellar dust we assume an isotropic and uniform distribution, extending from r = 0 to 30 au, based on 
the simulations of Grogan et al (1996).  We discuss below the expected departure from isotropy in a cone 
downstream of the Sun.
The outer cutoff for cometary and interstellar dust, corresponding to the inner edge of the Kuiper belt, is
arbitrary, and the fit to the IRAS data is not sensitive to this parameter.  The Grogan et (1996) estimate of
a negligible contribution by interstellar dust to the infrared foreground was based on a rather low density of interstellar
dust, 2x10$^{-27}$ gm cm$^{-3}$, at 5 a.u..  Kimura et al (2003) and Mann (2010) reports estimates 2-20 times higher 
than this and
we believe it is worth allowing the density of interstellar dust to be a free parameter and see whether
the resulting implied densities are unreasonable.

The interstellar dust grains can be expected to be smaller than those of cometary and asteroidal origin, and
are indeed found to be so in the spacecraft studies (eg Grun et al 1993).  Our best solution was found to be
with $Q_{\nu} \propto \nu$ for $\lambda > 24 \mu$m, =1 for $\lambda \leq 24 \mu$m, corresponding to
grain radius $\sim 2-4 \mu$m.  The corresponding temperature distribution for these smaller grains was taken to be
T = 305 $r^{-0.4}$ K.

Finally the $\sim$1 au ring and trailing 
blob are modelled using the formalism of Kelsall et al (1998), but we allow the amplitudes to be a
free parameter.

We estimated the rms fluctuation in each band at $|b|>40^o, |\beta|>20^o$ (after subtraction of the fan contribution)
to be $\sigma_{\nu}$ = 0.42, 0.55, 0.45 and 1.10 MJy sr$^{-1}$ at 12, 25, 60 and 100 $\mu$m
respectively.
We then minimise $\chi^2 =\Sigma [I_{\nu,IRAS} - C_{\nu} \int n(r,\lambda,\beta) B_{\nu}(T(r)) ds + D_{\nu}]^2/\sigma_{\nu}^2$.
Because IRAS did not have an absolute calibration there is uncertainty in the zero point in each band,
with a quoted rms 0.66, 0.73, 0.38, 1.19 MJy$sr^{-1}$ at 12, 25, 60 and 100 $\mu$m.  This makes the estimation
 of any isotropic component like our assumed interstellar component problematical.  

The approach we have followed here, is to use 
the offsets $D_{\nu}$ estimated for the IRAS data by the DIRBE team from direct comparison of DIRBE and IRAS data.  These are 
given as -0.48, -1.32, 0.13, -1.47 MJy $sr^{-1}$ at 12, 25, 60 and 100 $\mu$m (C.Beichman and S.Wheelock, 1993, 
IRAS project note, www.ipac.caltech.edu$/$ipac$/$newsletters$/$oct93$/$cobe.html).
We then mimimise the total $\chi^2$, summed over the four IRAS bands,
allowing a free calibration factor $C_{\nu}$.

Any isotropic component has to be consistent with the limits on an isotropic component set by DIRBE,
which are given as (2-$\sigma$) 1.9, 4.2, 1.5, 1.3 MJy$sr_{-1}$ at 12, 25, 60 and 100 $\mu$m (Hauser et al 1998) 
(see section 5). 

We find that both cometary and interstellar dust components were positively detected, in that the fit improved 
significantly when they were included.  This is consistent with the results of direct detection of interstellar dust by
Ulysses and other spacecraft (Grun et al 1993).

\section{Results of fits to IRAS data alone}
The results of varying the parameters of the model are shown in Table 1, with the penultimate column showing the 
sum of the reduced $\chi^2$ summed 
over the four IRAS bands.   The first line corresponds to the
preferred model of Jones and Rowan-Robinson (1993), with their parameters for the narrow and broad
bands.  The second line shows the effect of changing the radial density index to $\gamma$ = 1.3.

We now add in the ring and trailing blob, cometary and interstellar components, tune the amplitudes of these,
and then solve again for the parameters of the fan, Q and P.  The fit to the IRAS data is significantly improved.
We then make a number of small adjustments to the model which improve the fit: (i) we assumed that the outer 
narrow bands are due to the Veritas family, with
$\beta_{nb}$ = 9.35$^o$ and adjusted G2 to 0.12; (ii) we tuned the parameters for the symmetry plane of the
broad bands and found a better fit at $\Omega_{bb} = 110^o, i_{bb} = 2.6^o$; (iv) we tuned the parameters for the
symmetry plane of the fan and found $\Omega$ = 78$^o$, i = 1.50$^o$. 
A multi-parameter grid-search fitted to the IRAS data with parameters P, Q, z0, AMPBB, COM1, for different values of ISD1, yielded uncertainties of 2$\%$ for P, Q, and 10$\%$ for other parameters.  We estimate the uncertainty in ISD1 as 20$\%$. 
The best-fitting parameters are shown in line 3 of Table 1, shown in bold, model A.

\begin{figure*}
\epsfig{file=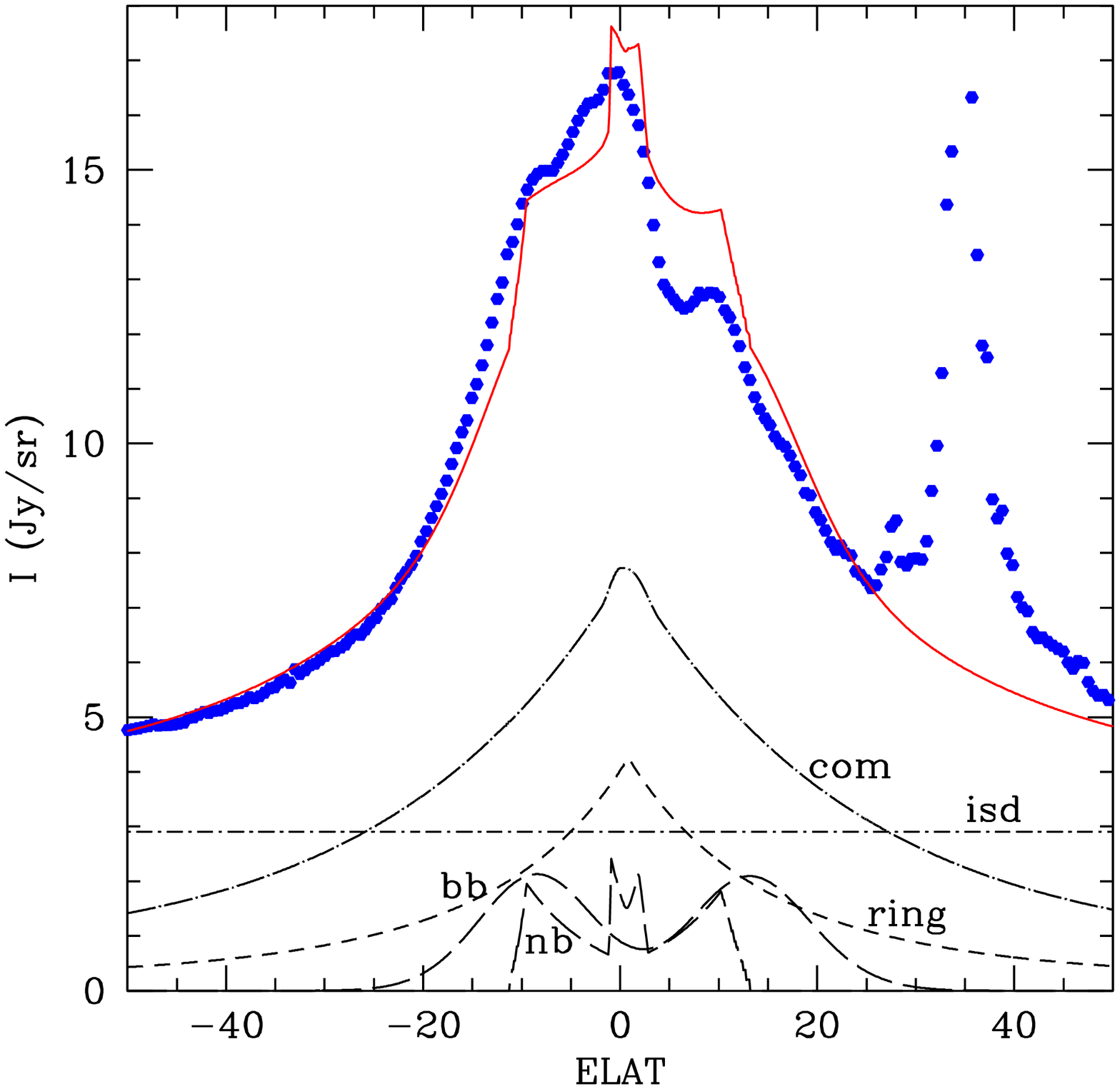,angle=0,width=7cm}
\epsfig{file=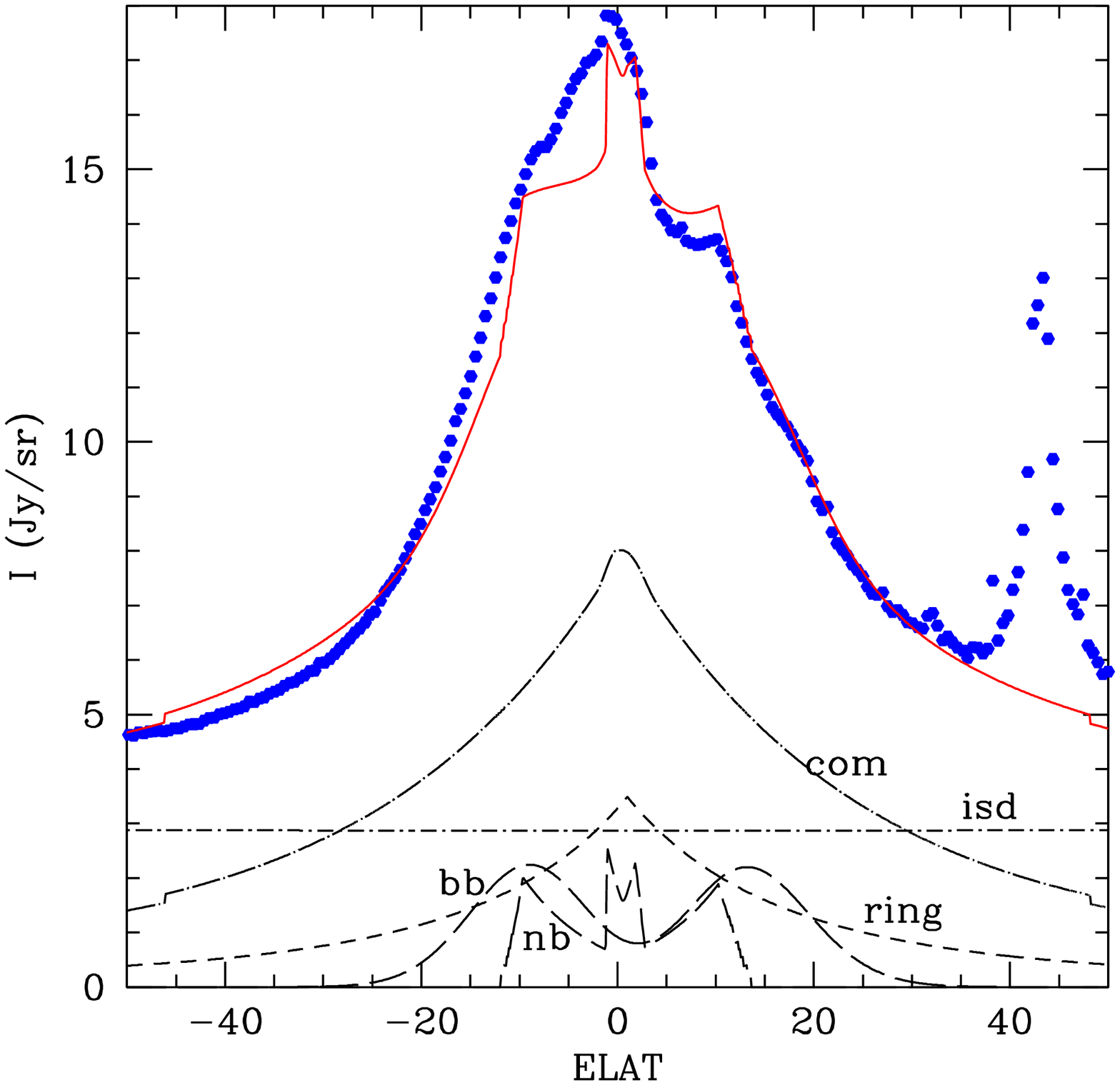,angle=0,width=7cm}
\epsfig{file=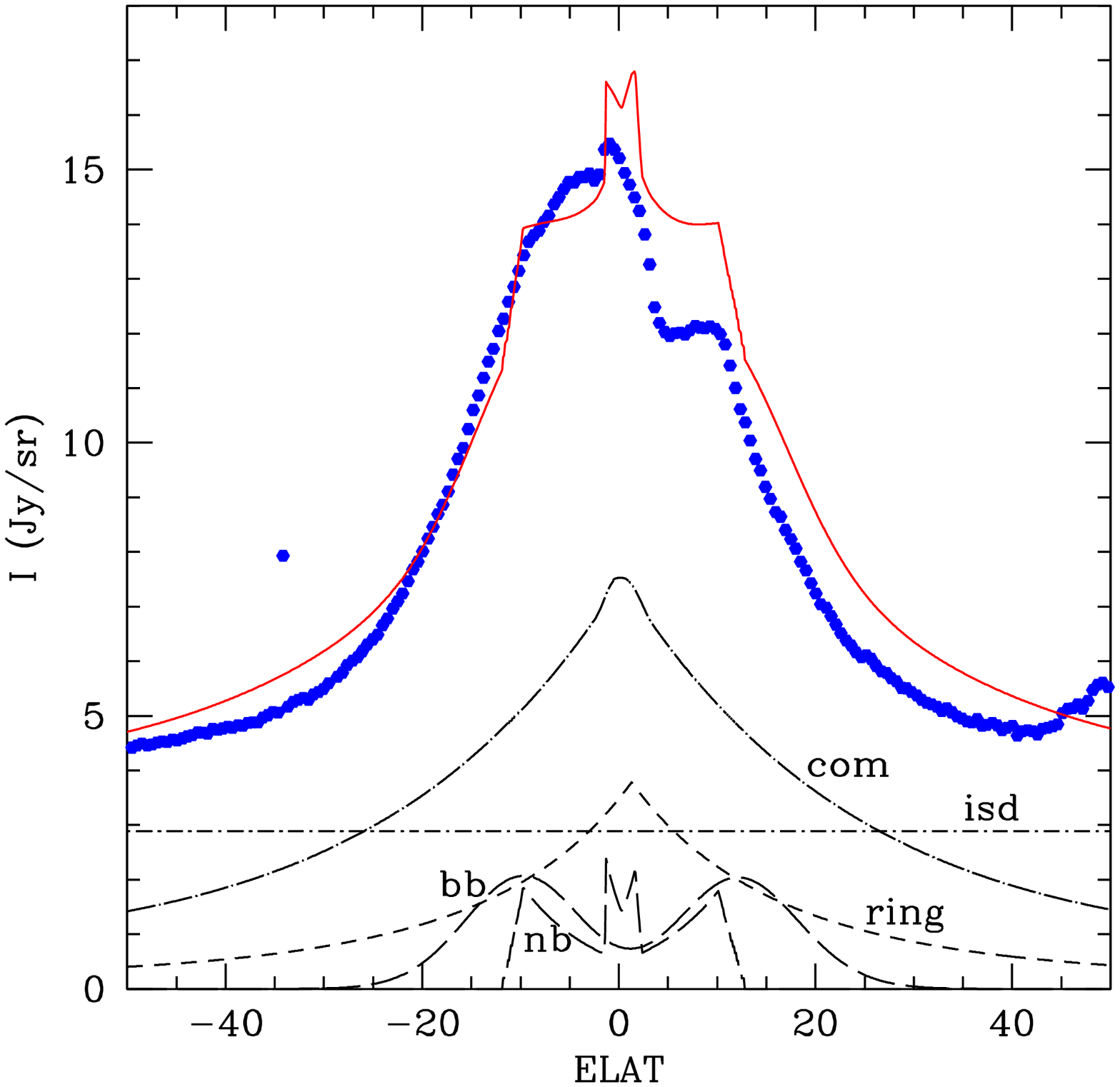,angle=0,width=7cm}
\epsfig{file=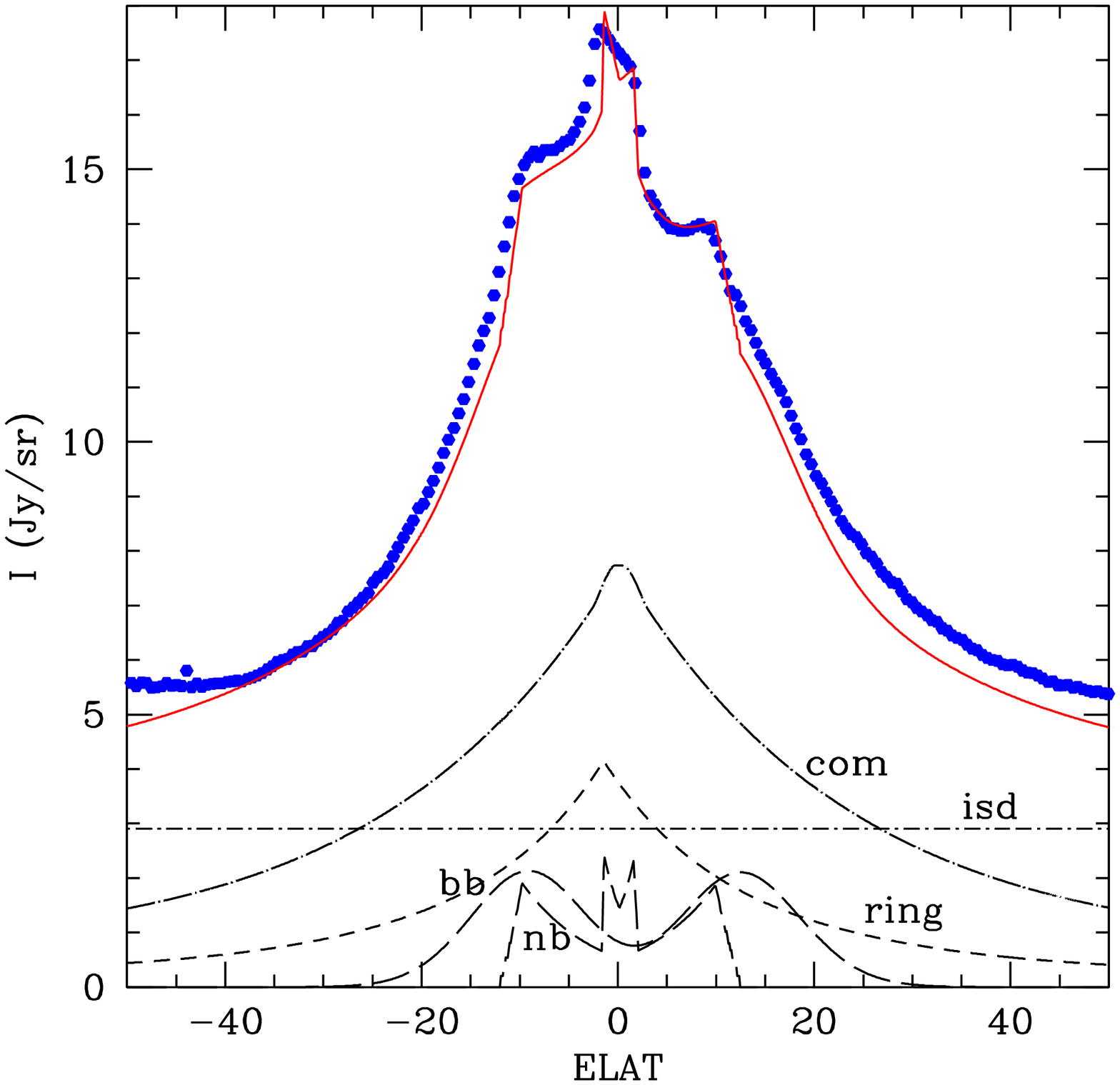,angle=0,width=7cm}
\caption{Comparison of IRAS scans at 25 $\mu$m, after subtraction of fan component (model A), with 
remaining components. Scans centred at (L to R) solar longitude 22.08, 33.40, 53.18 and 64.93$^o$.
}
\end{figure*}

\begin{figure*}
\epsfig{file=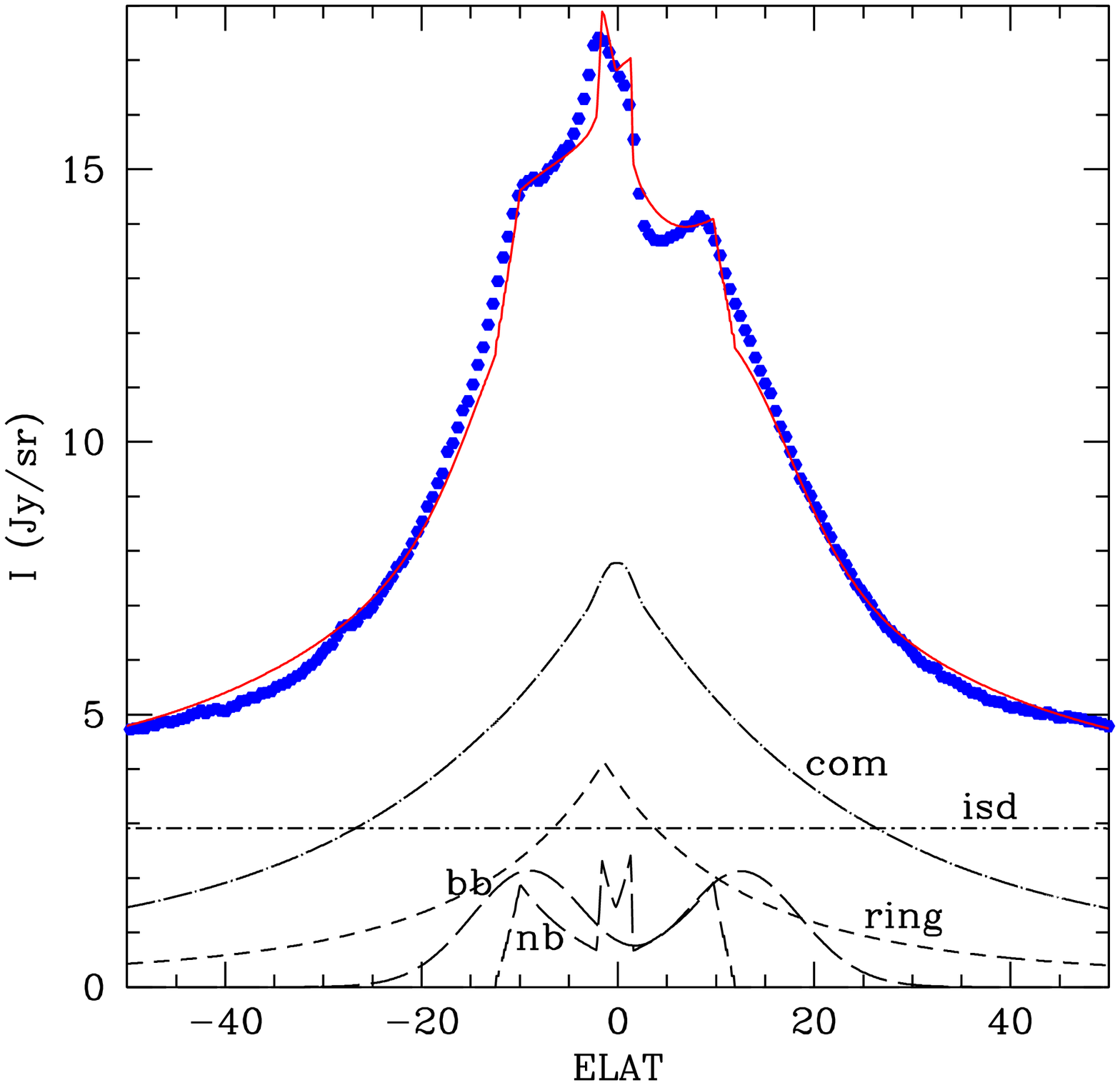,angle=0,width=7cm}
\epsfig{file=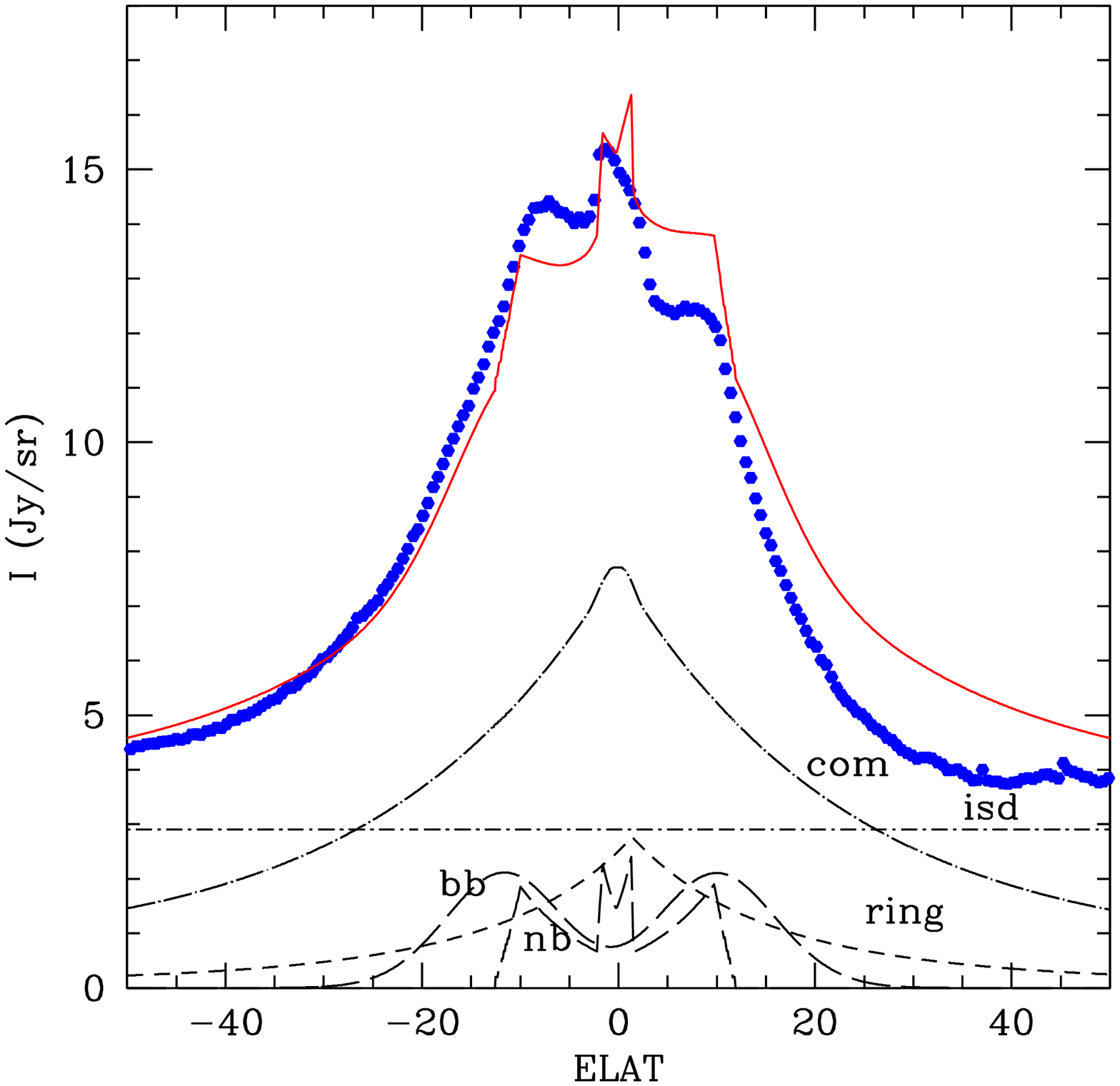,angle=0,width=7cm}
\epsfig{file=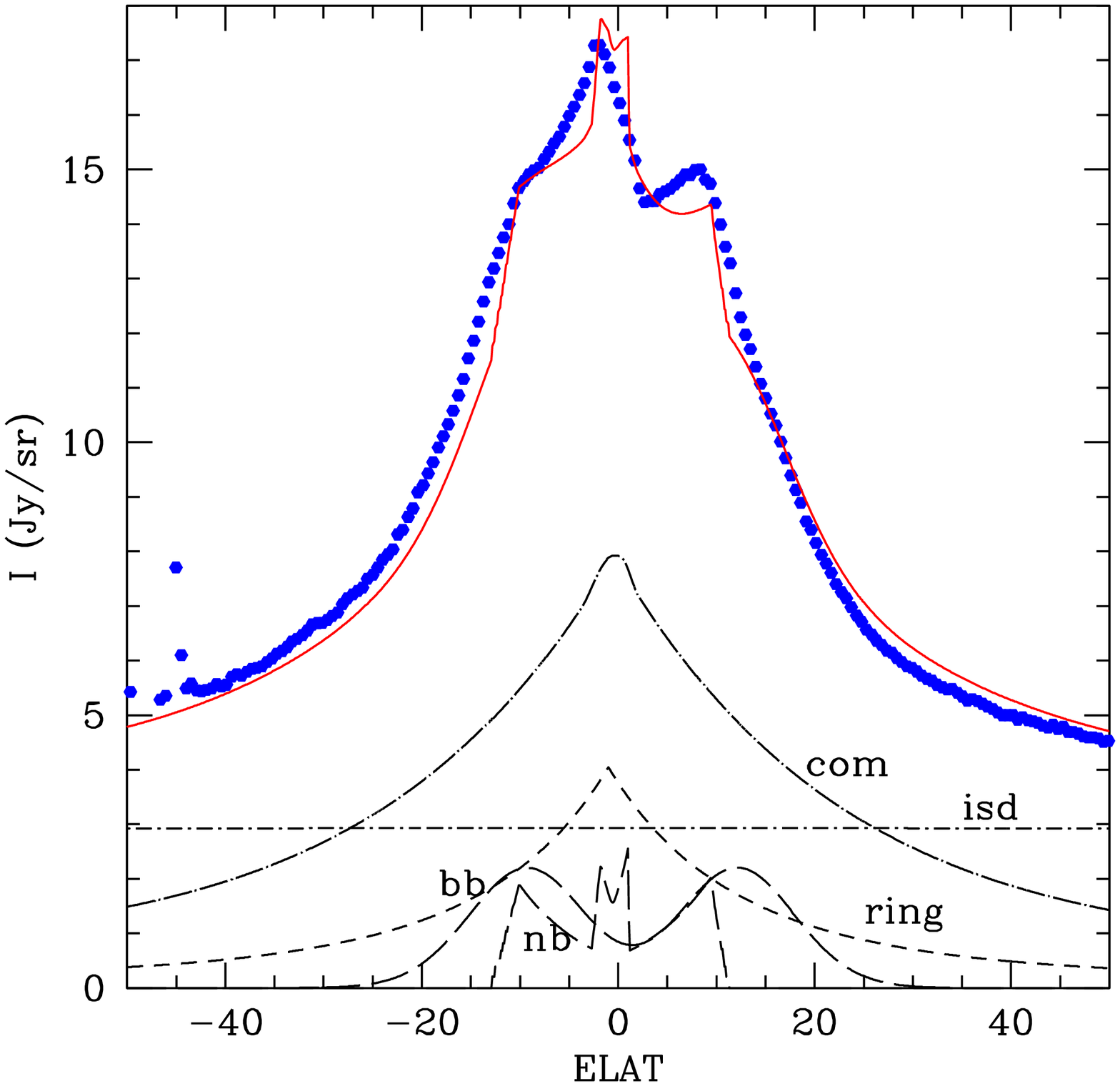,angle=0,width=7cm}
\epsfig{file=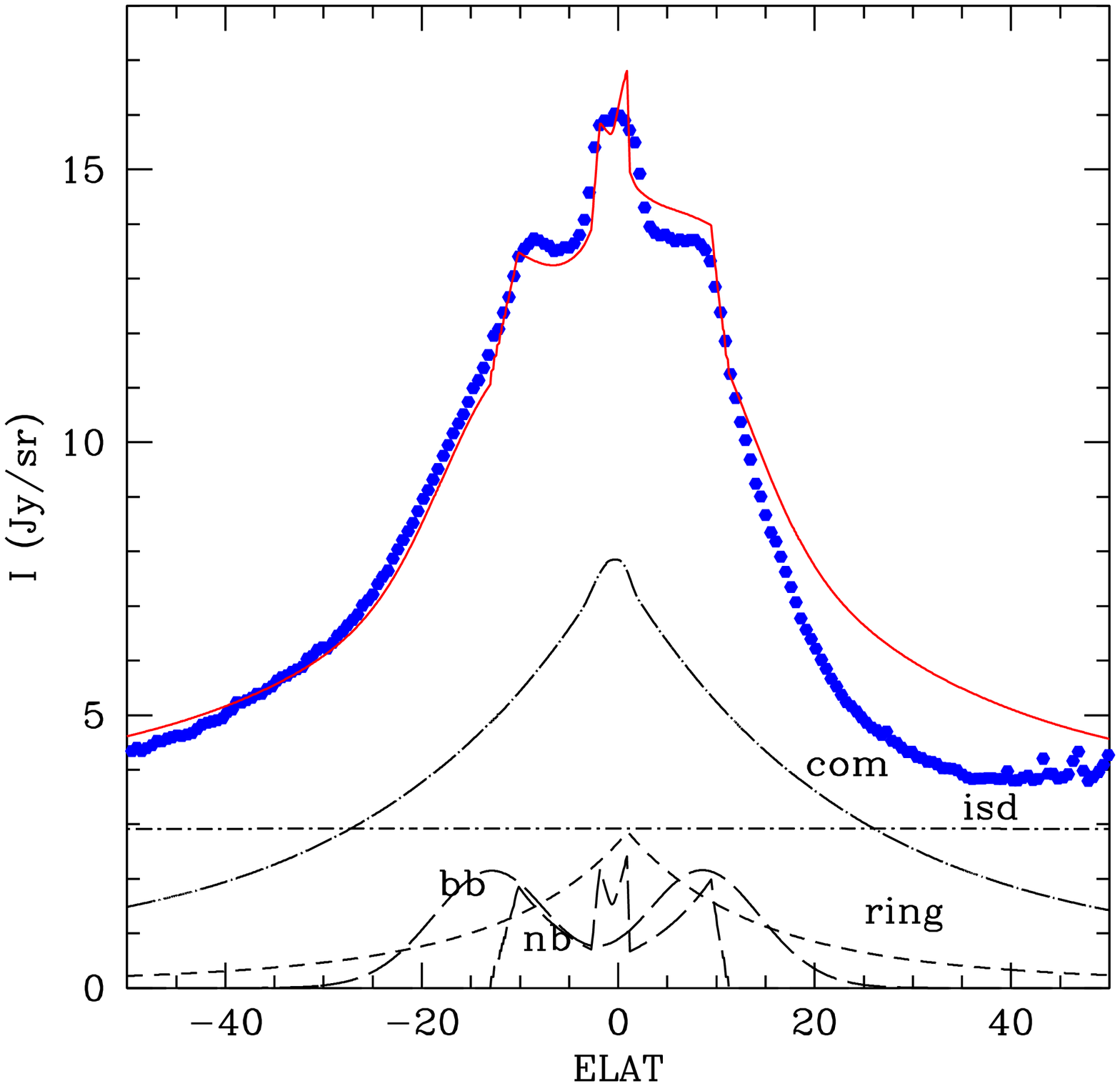,angle=0,width=7cm}
\caption{Comparison of IRAS scans at 25 $\mu$m, after subtraction of fan componen (model A)t, with 
remaining components. Scans centred at (L to R) solar longitude 90.61$^o$. 93.38, 121.61$^o$ and 124.99$^o$.
}
\end{figure*}

Figure 1 shows latitude profiles in the four IRAS bands for a scan with solar longitude 90.61$^o$, compared with the
predictions of model A.  The
fit is excellent.    Figures 2,3 shows latitude profiles at 25 $\mu$m for scans spread throughout the survey,
after subtraction of the fan component,
compared with the predictions of the other ingredients of the model.   The fits are spectacularly good.
The adjustments to the parameters of the narrow (asteroidal) and broad bands, and the inclusion of
components corresponding to cometary and interstellar dust, all contribute to the improvement of
the fits compared with the work of JRR.  We found the amplitude of the trailing blob needed to be smaller
than assumed by Kelsall et al (1998).  

For our basic model A, with the fan extending to 1.53 au, the asteroidal dust in the narrow and broad bands, 
cometary dust, and interstellar dust, integrated over all ecliptic latitudes, contribute 22.2, 70.4 and 7.5$\%$, respectively, of the density of dust,  relative to the fan.  The total  contribution from the three components adds to exactly 100$\%$ of the density in
the fan (this was not forced on the solution).   However it is unclear how much of the interstellar dust
within the Hoyle-Lyttleton column would make it into the fan.  Much could simply be channelled onto the
reverse side of the Sun.  Note that at the ecliptic plane in model A, the interstellar dust contributes only 1$\%$ of the
zodiacal dust density at 1 au.

\begin{figure*}
\epsfig{file=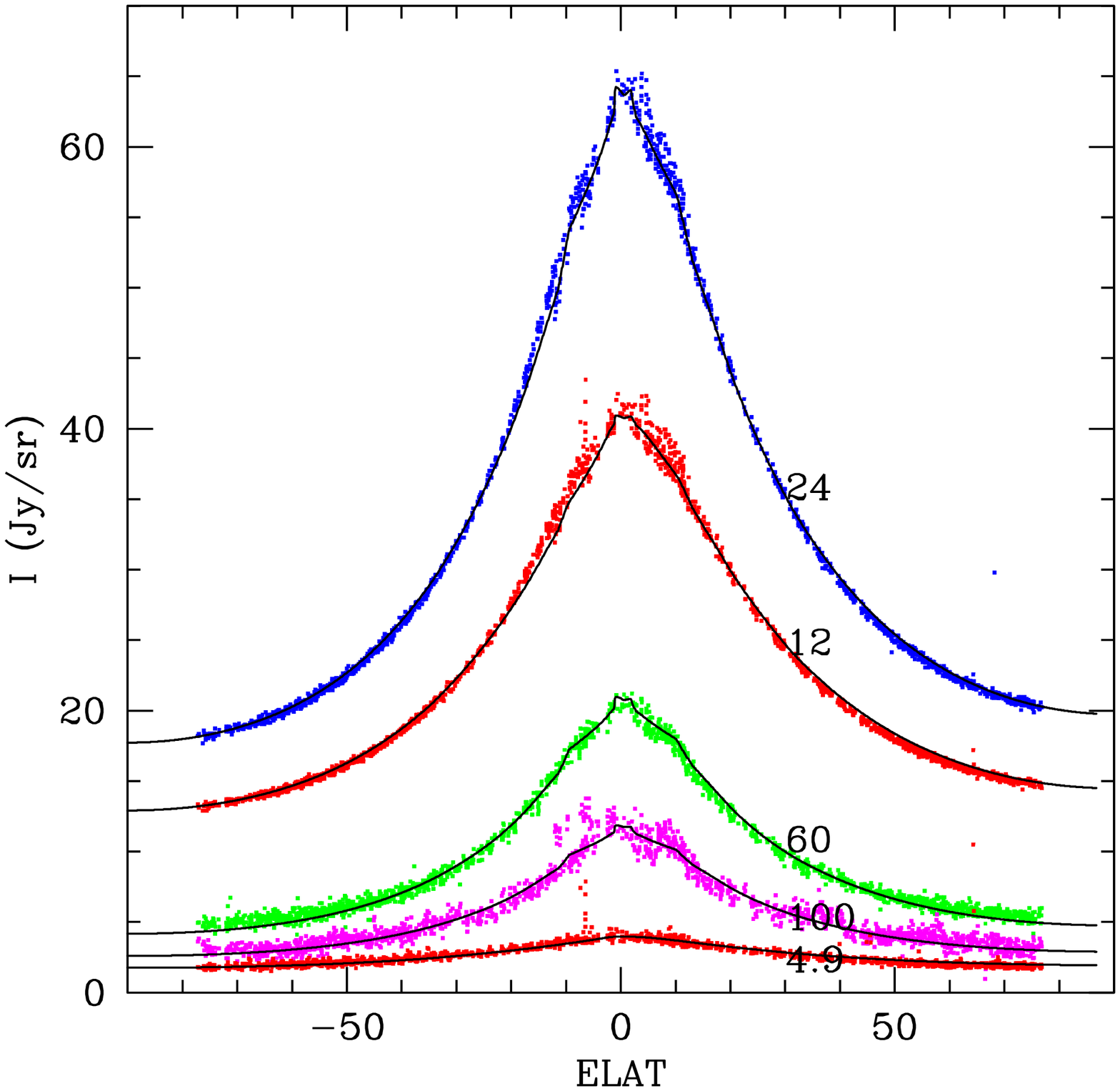,angle=0,width=10cm}
\caption{Comparison of DIRBE data at solar elongation 90$\pm1.0^o $, from Jan 19th 2009, restricted to $|b|>40^o$, 
at 4.9, 12, 25, 60 and 100 $\mu$m, with model A.
}
\end{figure*}

\begin{figure*}
\epsfig{file=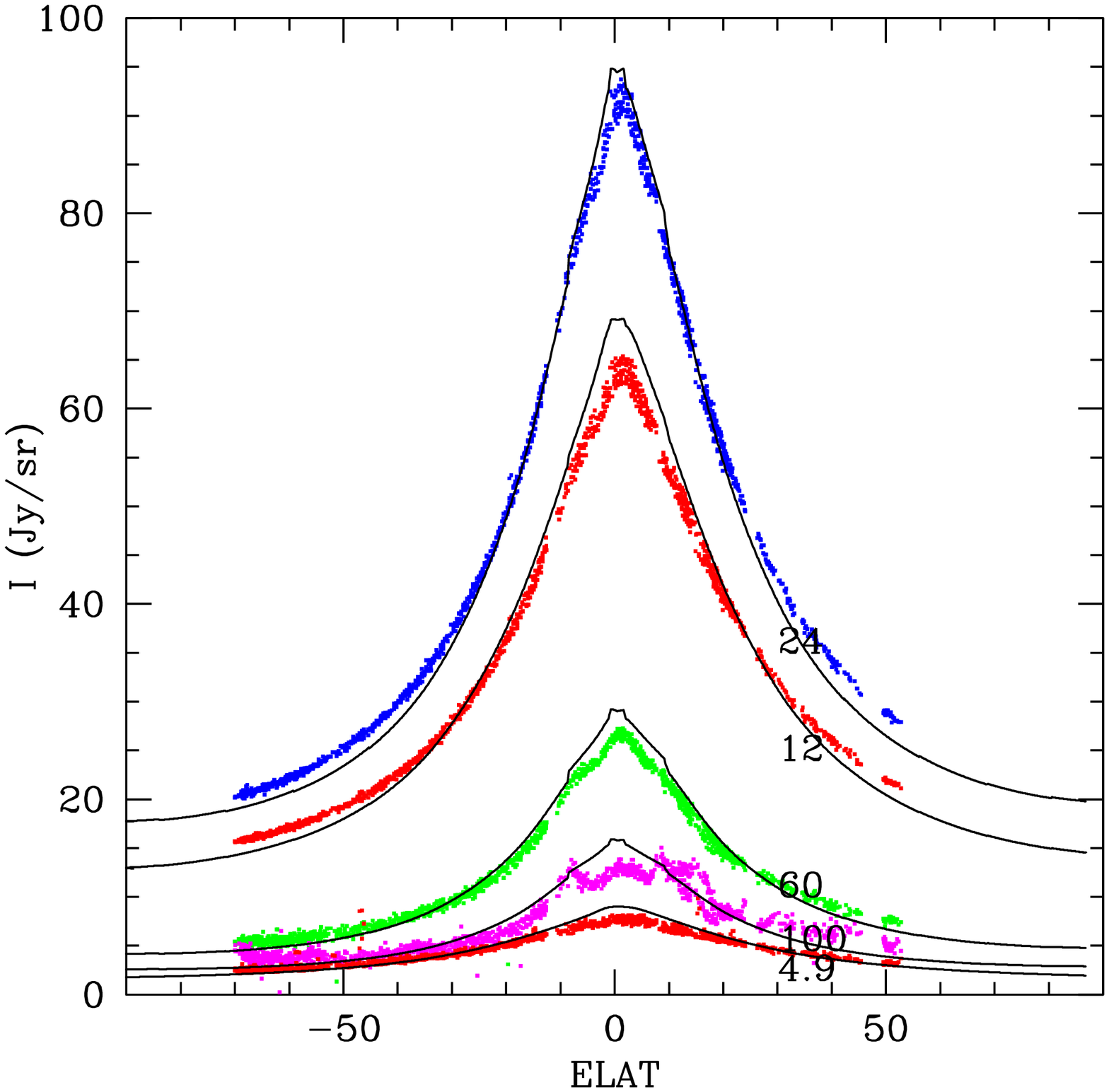,angle=0,width=7cm}
\epsfig{file=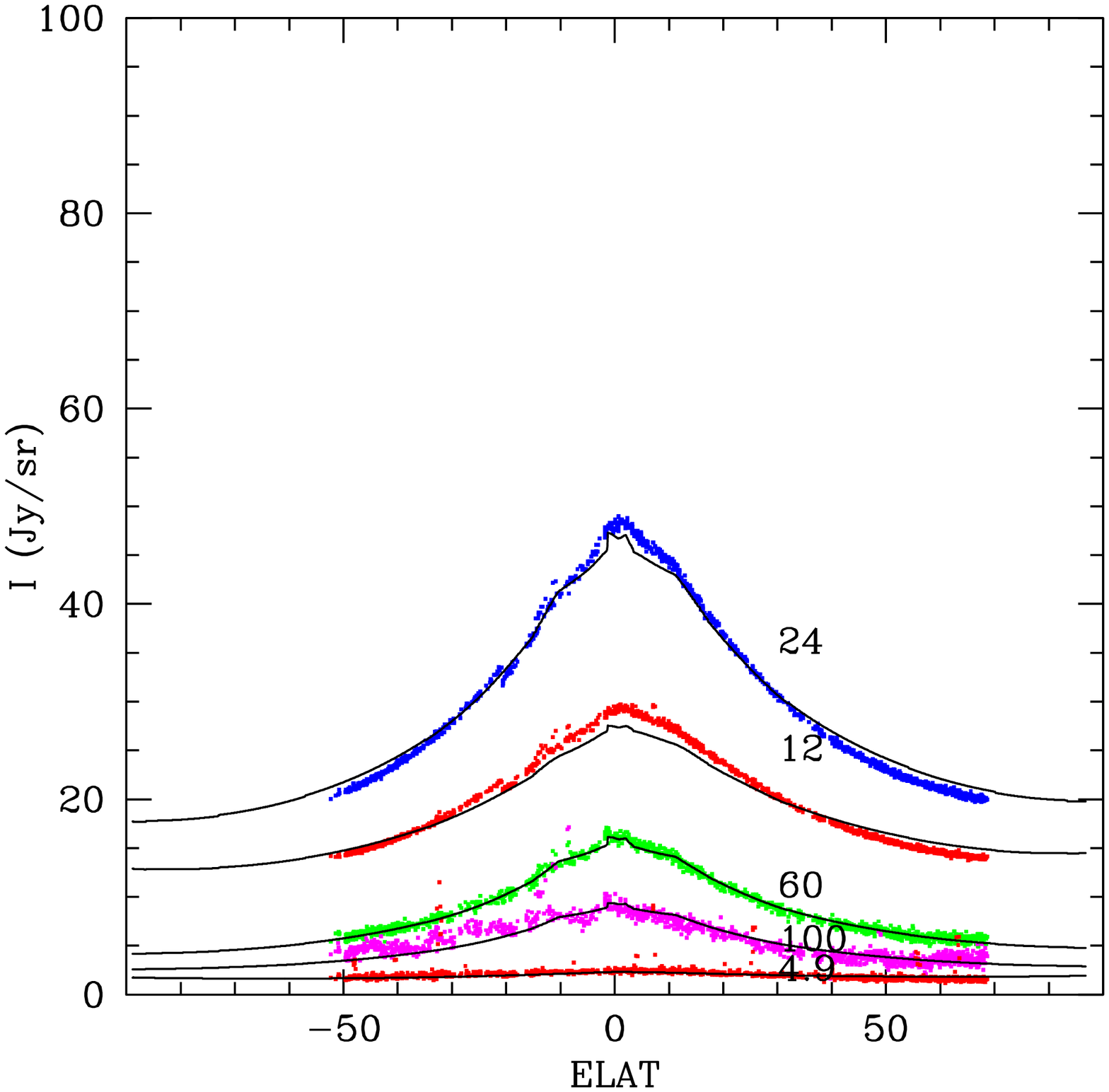,angle=0,width=7cm}
\caption{Comparison of DIRBE data at solar elongation 70$\pm1.0^o $ (L) and 110$\pm1.0^o $ (R), from Jan 19th 2009, 
restricted to $|b|>40^o$, at 4.9, 12, 25, 60 and 100 $\mu$m, with model A.
}
\end{figure*}

\section{Fits to DIRBE data}

We used the DIRBE calibrated data files, which are given for each day of the mission.  We first modelled 
the data for individual days using our best model A from Table 2, excluding days where $\Sigma rms^2 > 5 ($MJy sr$^{-1})^2$, 
which can arise for example because the Moon
is in the field of view during the day.  Data were used only if there were detections in all four bands at 12, 25, 60 and 
100 $\mu$m, and only sky at $|b| > 40^o$ was used in the zodiacal modelling.  We focussed on three blocks of data 
where there were a significant number of contiguous days
with good data, at day numbers 19-33, 51-95 and 216-264 in 1990.  40 of the 109 days in these three blocks
were excluded.  For a further 54 days analysed, distributed at random through the mission, none satisfied
our constraint.  The 69 days of data used in our solution covered 92$\%$ of the sky, due to the particular scan
strategy of COBE.  After exclusion of $|b| < 40^o$ and areas affected by cirrus, we used 1.5 million observations in the solution,
representing about 3$\%$ of the total data.  This is more than a factor of 20 times the amount of
data used by Kelsall et al (1998) in their solution (0.13$\%$ of the total data).  We made no restriction on
solar elongation angle.

We estimated the rms fluctuations for the DIRBE data at $|b|>40^o, |\beta>20^o$, after subtraction of the model A fan, as 0.80, 0.93, 0.67 and 0.80 MJy sr$^{-1}$ at 12, 25, 60,
100 $\mu$m, and calculated $\chi^2$ as above (last column in Table 1).

Figure 4 shows fits with models A to data at solar elongation 90$\pm 1.0^o$, $|b|>40^o$, on Jan 19th,
1990.  The nature of the COBE scan strategy do not allow pole-to-pole scans to be plotted, as for IRAS.  We have shown
fits to the 4.9 $\mu$m data, but have not used the latter in the parameter fits.  The data at 140 and 240 $\mu$m
were too noisy to plot, or to use in the solution.  The model fits to the COBE data are excellent, and by excluding directions at
$|b|<40^o$ from the plots, we achieve much more compelling comparisons of the models with the DIRBE data
than shown by Kelsall et al (1998) (their Fig 8).  The combination of the larger beam of DIRBE, the precessing scan strategy
of COBE, and the poorer signal-to-noise compared with the IRAS scans, means that details of the asteroidal dust
bands are much more poorly resolved by DIRBE.  We were also unable to detect the trailing blob in the DIRBE data so
the amplitude of this component could be determined only from IRAS data.  
 Figure 5 show the corresponding plots for solar elongation 70 and 110$\pm 1.0^o$, illustrating that the 
model works well over a wide range of solar elongation.

We should check that the magnitude of our claimed isotropic interstellar dust component is not
inconsistent with limits set in earlier studies.  The magnitude of the isotropic component due to interstellar dust
at 12, 25, 60, and 100 $\mu$m, respectively, are for model (A) 1.59, 2.90, 0.68, 0.23 MJy sr$^{-1}$.  These are
consistent with the limits on an isotropic background set by Hauser et al (1998).  
Figure 6, which shows $\chi^2_n$ as a function of the amplitude in the interstellar dust component, keeping the
other parameters of model A fixed,  illustrates the strong 
detection of the interstellar dust component in both IRAS and DIRBE data.  

\begin{figure}
\epsfig{file=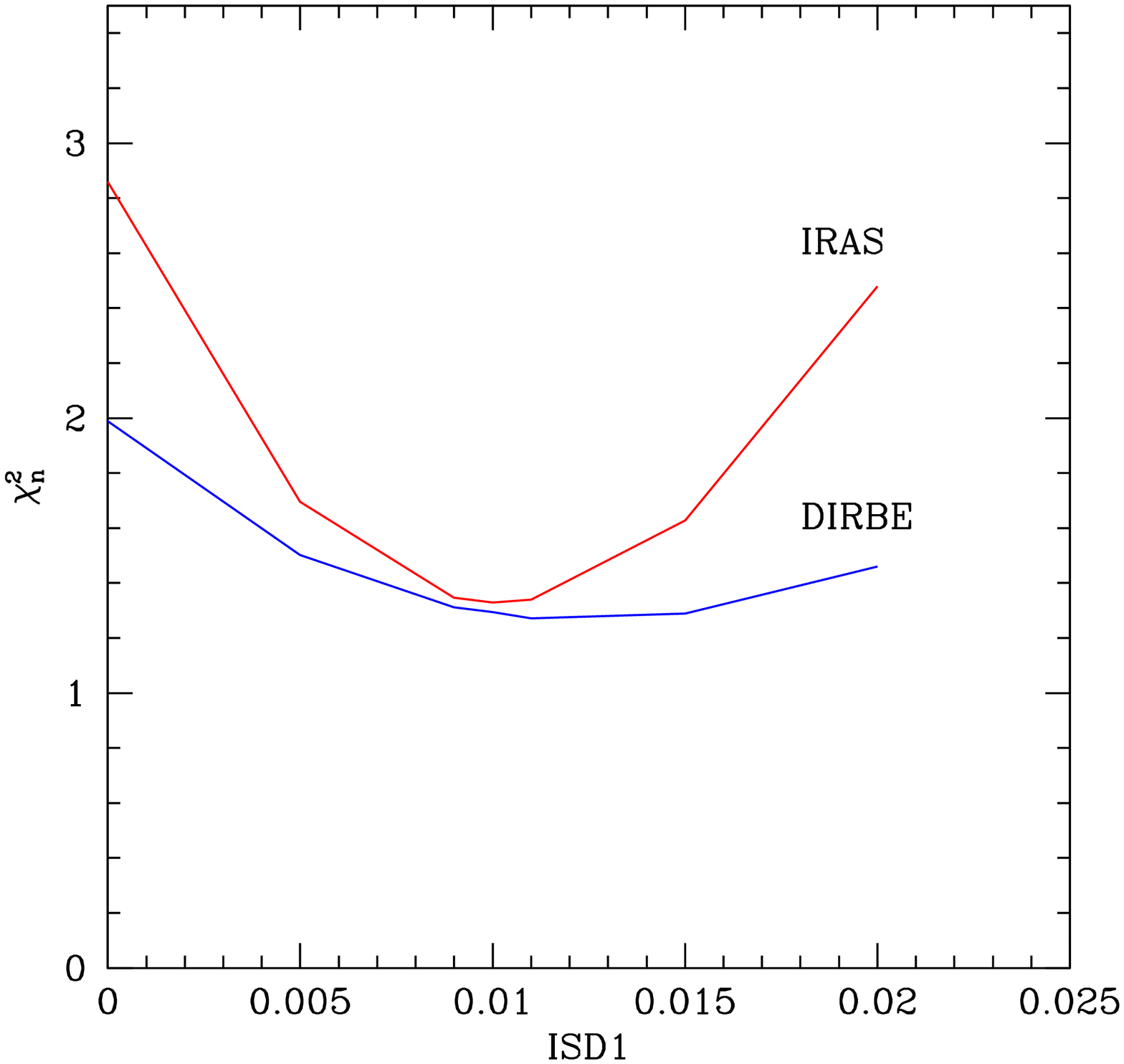,angle=0,width=7cm}
\caption{$\chi^2_n$ versus the interstellar dust amplitude, ISD1, for IRAS and DIRBE data.
}
\end{figure}

For model A we investigated whether extending the broad bands to 30 au (as proposed by Rowan-Robinson et al 1991) affected the fit.   
We found no significant improvement in the $\chi^2_n$ for either the IRAS and DIRBE data.   We also investigated
whether making the broad band dust grains smaller improved the fit, but again there was no improvement.
However the origin of this broad-band component does merit further study.  It may arise from an older collision event
between asteroid family members in the main belt.  A much more distant origin in the Kuiper belt can not be ruled out.


\begin{figure}
\epsfig{file=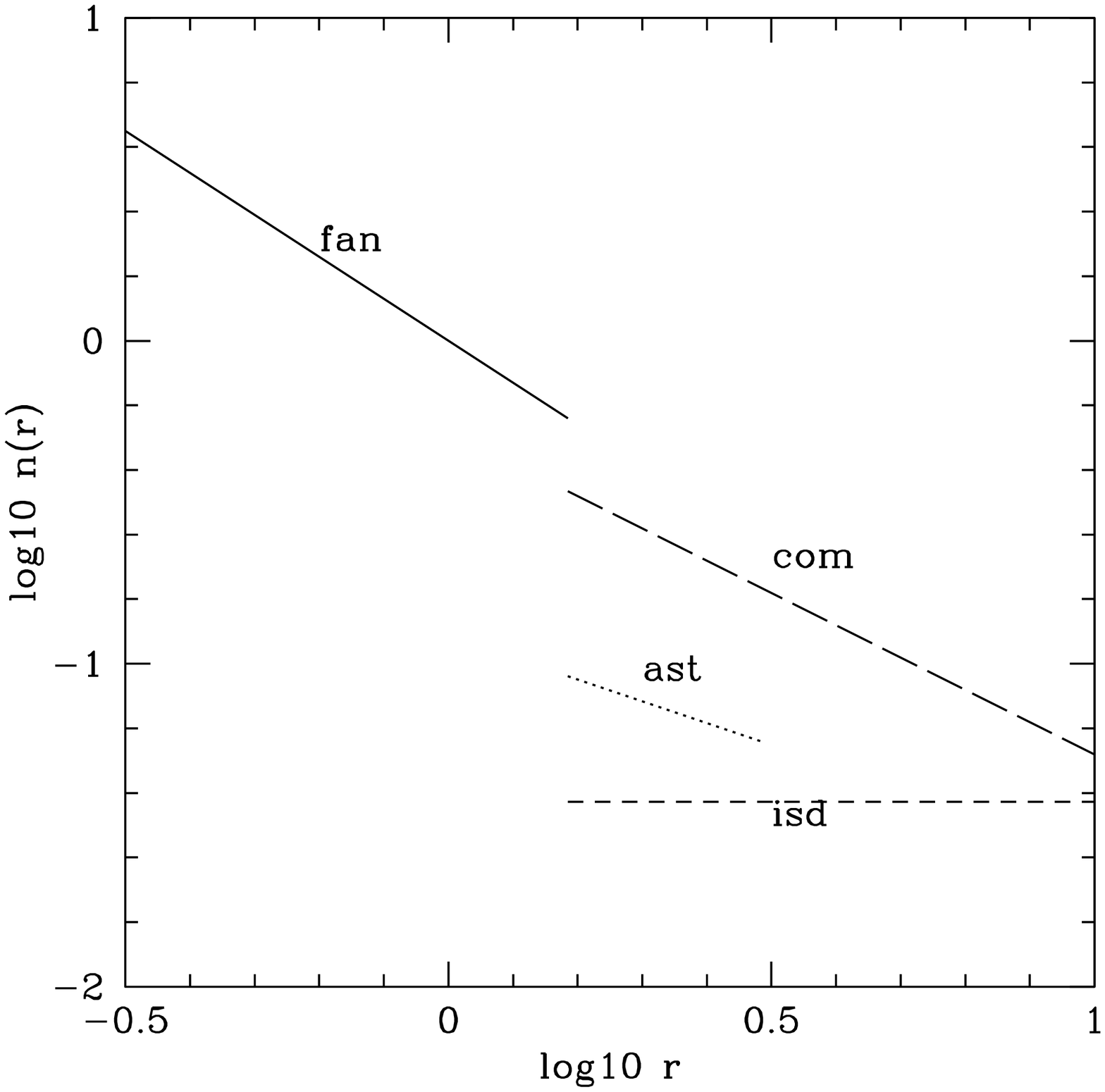,angle=0,width=7cm}
\caption{Dust density versus radius (a.u.) for fan, cometary, asteroidal and interstellar dust in model A.
}
\end{figure}

\section{Models with fan extended to 5.2 a.u.}

Although the direct evidence from spacecraft data is consistent with the fan extending to r = 1.53 au, we have explored
the possibility that it could extend to the orbit of Jupiter at r = 5.2 au.  We assume that 15 $\%$ of the
zodiacal cloud interior to 1.53 au is supplied by the asteroidal bands, so that the amplitude
of the extension from r =1.53 to 5.2 au is at an amplitude 0.85 times that interior to 1.53 au.
The main effect of such an extension is to
increase the intensity at $|\beta| < 30^o$.   For our modified fan (eqn (3)) to remain consistent with observations it is
necessary to change the parameters P, Q, but also to increase $z_0$.  Line 5 of the Table shows a fit
with Q=3.5, P=2.5 and $z_0$=0.15 (model B, see Fig 8).  It was also necessary to reduce the amplitude of the outer narrow bands, AMP2, 
and of the broad bands, AMPBB.  The best combined fit to IRAS and DIRBE data has no cometary dust beyond 5.2 a.u., 
and the amplitude of interstellar
dust is lower by a factor of 3 than in model A.  The fits to the IRAS data are significantly worse than model A.
Figure 7L shows the fit to the IRAS data at 25 $\mu$m after subtraction of the fan component.

\begin{figure*}
\epsfig{file=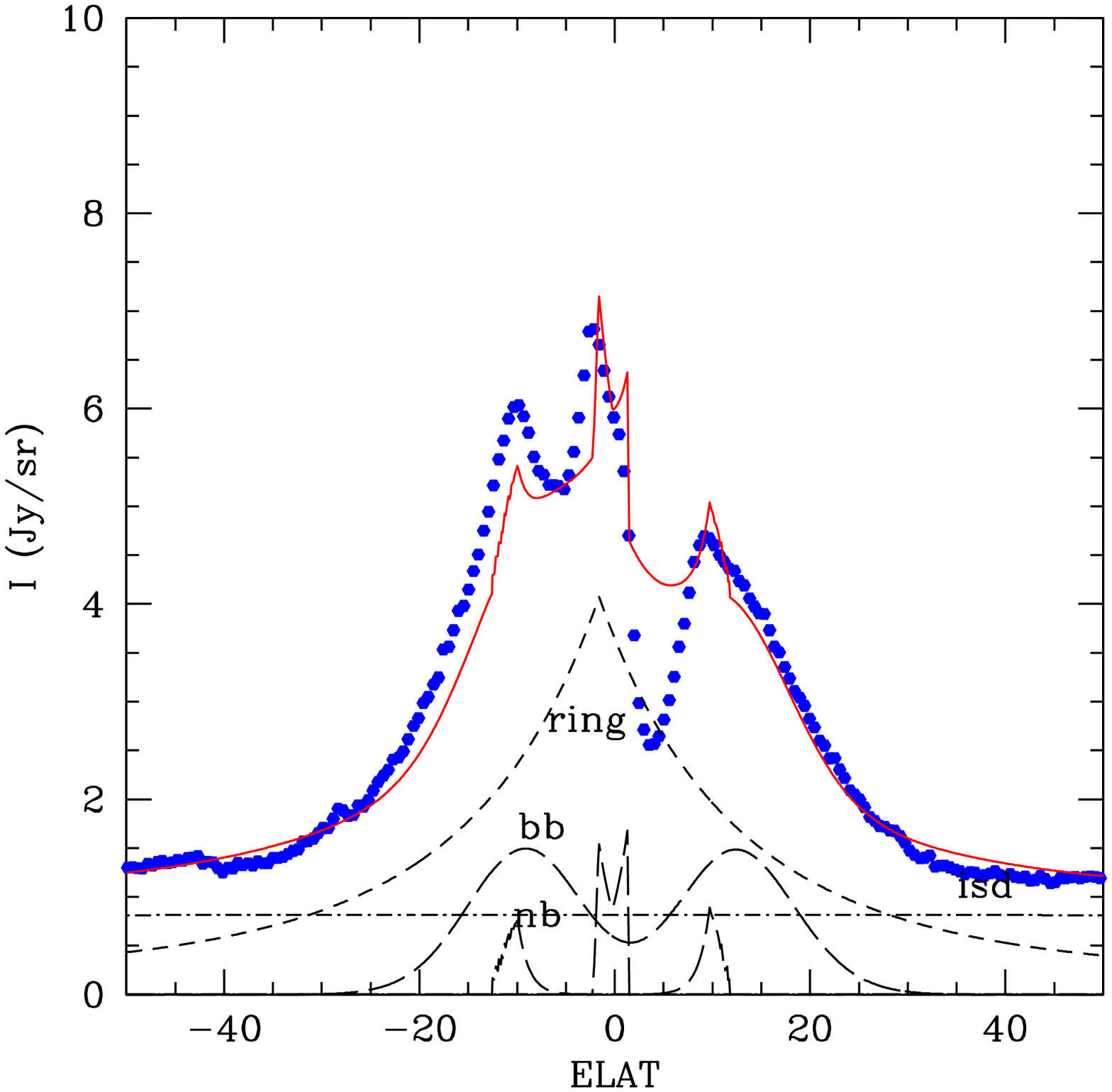,angle=0,width=7cm}
\epsfig{file=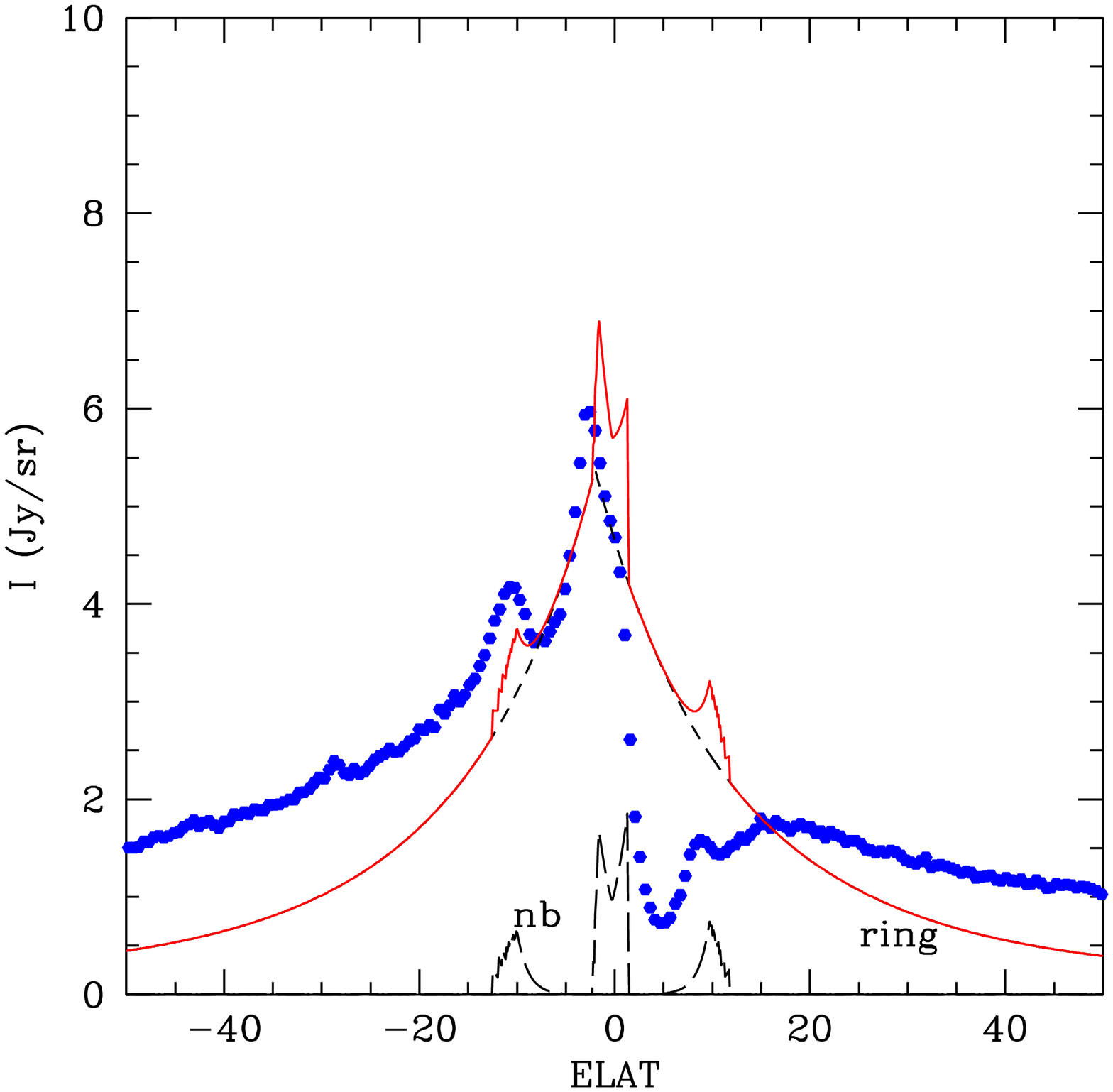,angle=0,width=7cm}
\caption{Comparison of IRAS scans at 25 $\mu$m, (L) after subtraction of fan component, extended to 5.2 au (model B), with 
remaining components; (B) after subtraction of Kelsall et al (1998) fan (model K) . Scans centred at solar longitude 90.61$^o$.
}
\end{figure*}

We have also fitted the IRAS data with the Kelsall et
al (1998) fan model, plus ring and trailing blob.  For the narrow bands we have simply used eqn (4), with adjustments
to the amplitudes and to G2 (=1.0).  The Kelsall et al fan changes shape
at $\beta_0 \sim$ 10$^o$ and thereby dispenses with the need for the broad bands.  The fit is noticeably worse for the
IRAS data, with $\chi^2_n$ = 3.40 ($D_{\nu}$ = 0).
The extension of the fan to 5.2 au is not supported by the Divine (1993) analysis.



\section{Mass-density in zodiacal dust}

The parameter $C_{\nu}$ can be interpreted as $\pi a^2 Q_{\nu} n_0 a_0$, where $a, Q_{\nu}$ are the characteristic
grain radius and absorption efficiency, $n_0$ is the grain number-density and $a_0$ = 1 au.  Hence we can
derive an order-of magnitude mass-density in grains, $\rho_{gr} = n_0 (4 \pi a^3 \rho_0/3)$, where $\rho_0$ is the mean density 
of the grain, taken to be 2.5 gm cm$^{-3}$.

For model A,  in the ecliptic plane at r = 1.53 au, and using $C_{\nu}$ at $\lambda = 12 \mu$m,we find\\
\medskip
  $\rho_{gr} (Q_{\nu}/a(\mu m)) \sim10^{-23.27}$ gm cm$^{-3}$, for the fan\\
\indent	$ \sim 10^{-25.27}$ gm cm$^{-3}$, for the interstellar dust\\
\indent     $ \sim10^{-23.67}$ gm cm$^{-3}$, for the cometary dust.\\

For comparison Kimura et al (2003) estimate the mass-density of interstellar dust at r = 1.5 au
as  $10^{-25.3 \pm 0.7}$ gm cm$^{-3}$, so our value is consistent with this for a $\sim 2-4 \mu m, Q_{\nu} \sim 1$.
On the other hand they estimate the mass-density of interstellar dust at r $>$ 4 au as $10^{-26.43 (+0.26, -0.60)} gm cm^{-3}$, attributing the difference to gravitational focussing of the dust in the inner solar
system.  Our assumed isotropic uniform dust density is therefore an order of magnitude higher than
that measured by Ulysses at r $>$ 4 au.  Further work will be needed to assess whether the density of interstellar 
grains has a strong dependence on radial distance.  Our dust density estimates are approximate, but the satellite
estimates are also indirect, based on the measured charge.

\begin{table*}
\caption{Parameters for new zodiacal dust models}
\begin{tabular}{llllllllllllllll}
fan  & & & &  narrow & bands & broad b. & ring & blob & comet & & i.s.d. & $\chi^2_n$ &  \\
Q & P & z0 & RMAX  & AMP1 & AMP2 & AMPBB  & AMPSR & AMPBL & COM1 & PCOM & ISD1 & IRAS & DIRBE \\
 &&&&&&&&&&&&&&\\
 8.0 & 2.85 & 0.065 & 1.52 & 0.030 & 0.039 & 0.029  & - & - & - & - & - & 5.03 &($\gamma$=1.0)\\
 8.0 & 2.85 & 0.065 & 1.52 & 0.030 & 0.039 & 0.029  & - & - & - & - & - & 4.32 &($\gamma$=1.3)\\
 \bf{10.7} & \bf{2.13} & \bf{0.06} & \bf{1.52}  & \bf{0.032} & \bf{0.040} &\bf{ 0.051}  & \bf{0.16} & \bf{0.065} & \bf{0.37} & \bf{2.5} &\bf{0.010} & \bf{1.33} & \bf{1.30 A}\\
 &&&&&&&&&&&&&&\\
3.5 & 2.5 &  0.15  & 5.2 & 0.035 & 0.035 & 0.030  & 0.17 & 0.06 & 0.0 & - & 0.003 & 2.48 & 1.33 \bf{B}\\
 &&&&&&&&&&&&&&\\
0.0 & 4.14 & 0.19 & 5.2 & 0.035 & 0.027 & 0.0  & 0.17 & 0.16 & 0.0 & - & 0.0 & 3.40 & 1.31 \bf{K}\\
&&&&&&&&&&&&&&&\\
\end{tabular}
\end{table*}

\section{Discussion}

Our model provides fits to the IRAS and DIRBE data significantly better than those achieved in previous
analyses and shows good consistency with the major components identified
by Divine (1993) in the Ulysses spacecraft data.  The major new constituents compared to JRR 
are the interstellar and cometary components, corresponding to the 'halo' component of Divine (1993).  

In model A, the relative contributions of cometary, interstellar and asteroidal dust to the density of the fan at
1.5 au are 70.4$\%$, 7.5$\%$ and 22.2$\%$, and the density of the fan is exactly accounted for by these components.
This is the first analytical fit to the zodiacal infrared data in which the origin of the fan is fully accounted for.
Our assumed isotropic uniform density of interstellar grains is consistent with values measured by
Ulysses at 1.5 au, but an order of magnitude higher than values measured at r $>$ 4 au.  Further work will be needed 
to assess whether the density of interstellar grains has a strong dependence on radial distance, as proposed
by Kimaru et al (2003).

Our approach gives a valid independent perspective on dust in the solar system to that given by dynamical 
simulation of the different components (eg Gustafsen et al 1987, Durda et al 1997, Dikarev et al 2005,
Ipatov et al 2008, Nesvorny et al 2010).  Although predictions of dynamical models with just asteroidal and 
cometary dust have been compared with IRAS (Nesvorny et al 2010) and COBE (Dikarev et al 2005) data, 
we find that to fit both sets of data to the accuracy achieved here requires the additional ingredient of a 
homogenous, isotropic component, corresponding to interstellar dust.

While cometary and asteroidal dust spiral inwards very slowly and can experience strong orbital changes at
the crossing of planetary orbits, the fast-moving interstellar dust is almost unaffected by the planets and is trapped 
in the inner solar system by the sun's gravity through Hoyle-Lyttleton accretion.  The smallest grains will also be
repelled from a cylindrical column behind the Sun by radiation pressure and magnetic forces.  This column
would also show a concentration of larger grains being accreted towards the Sun.  Unfortunately
the direction of this column ($(\lambda,\beta) \sim (-172^o, -5^o)$, Kimaru et al 2003) is rather close to the ecliptic plane, 
so it can not be resolved in maps of the IRAS background emission.  Detection of the downstream accretion column
would be an important confirmation of the interstellar dust component.

Nesvorny et al (2010) carried out an interesting simulation of the dynamics of cometary dust in the solar system.
Their estimate that cometary dust provides over 90$\%$ of the zodiacal dust in the inner solar system
is higher than our model predicts.  We estimate that cometary dust could contribute 
60-80$\%$, with asteroidal and interstellar dust contributing 20-40$\%$
between them.  Our estimate of the relative proportions of cometary and asteroidal dust agree well with the
estimate of Liou et al (1995) from a dynamical model.

We find the contribution of interstellar dust may be more significant than estimated by Grogan et al (1996).  The DIRBE
data is crucial here in demonstrating the presence of an isotropic foreground component at 12, 25, 60 and 100
microns.  The amplitude of this foreground does not conflict  with the isotropic background limits set 
by Hauser et al (1998).

We find strong support for the Hicks, May and Reay (1974) proposal, extended by May (2007), based on kinematical 
observations, that interstellar dust is a significant  contributor to the local zodiacal dust cloud,
although the density we find in the ecliptic plane is much lower than that estimated by May.
There is a need for a new ground-based kinematic study of zodiacal dust, ideally extended over several years
to test for time variation through the solar cycle.  The models presented here can also be tested by dynamical 
simulations and by further in situ measurements from spacecraft at r $>$ 5 au.

The zodiacal dust cloud appears to be supplied by a combination of cometary dust, dust from collisions 
between members of asteroid families in the asteroid belt, and interstellar dust.  It therefore carries detailed 
information about the recent history of the Sun's debris disk and merits far more intensive study than it has 
received to date.


\end{document}

%% file: macro.tex
\newcommand{\aip}{{\small ${\cal AIPS}$}}
\newcommand{\gtsim}{\mbox{{\raisebox{-0.4ex}{$\stackrel{>}{{\scriptstyle\sim}}
$}}}}
\newcommand{\ltsim}{\mbox{{\raisebox{-0.4ex}{$\stackrel{<}{{\scriptstyle\sim}}
$}}}}
\newcommand{\s}{$\stackrel{\rm s}{.}$}
\newcommand{\h}{$^{\rm h}$}
\newcommand{\m}{$^{\rm m}$}
\newcommand{\pp}{$\stackrel{\prime\prime}{.}$}
\newcommand{\de}{$^{\circ}$}
\newcommand{\p}{$^{\prime}$}
\newcommand{\arc}{$^{\prime\prime}$}
\newcommand{\marc}{^{\prime\prime}}
\newcommand{\rs}{{\em $r_s$}}
\newcommand{\DPM}{{\em DPM}}
\newcommand{\alf}{{\displaystyle\biggl({\nu_{\rm h} \over \nu_{\rm l}}\biggr)^{\alpha}} }

\newcommand{\figstart}[1]
    { \begin{figure}[htb]
      \begin{picture}(0,#1) }
\newcommand{\figend}[4]
    { \end{picture}
      \special{#1}
      \caption[#2]{#3}
      \label{#4}
      \end{figure} }
\newcommand{\fig}[5]
    { \figstart{#1}
      \figend{#2}{#3}{#4}{#5} }
\newcommand{\bHS}{\beta_{\mbox{\scriptsize HS}}}
\newcommand{\bBF}{\beta_{\mbox{\scriptsize BF}}}
\newcommand{\nT}{\nu_{\mbox{\scriptsize T}}}
\newcommand{\et}{E_{\mbox{\scriptsize T}}}
\newcommand{\nTn}{\nu_{\mbox{\scriptsize Tn}}}
\newcommand{\nTf}{\nu_{\mbox{\scriptsize Tf}}}
\newcommand{\tn}{\tau_{x\mbox{\scriptsize n}}}
\newcommand{\tf}{\tau_{x\mbox{\scriptsize f}}}
\newcommand{\xn}{x_{\mbox{\scriptsize n}}}
\newcommand{\xf}{x_{\mbox{\scriptsize f}}}
\newcommand{\yn}{y_{\mbox{\scriptsize n}}}
\newcommand{\yf}{y_{\mbox{\scriptsize f}}}
\newcommand{\lln}{l_{\mbox{\scriptsize n}}}
\newcommand{\llf}{l_{\mbox{\scriptsize f}}}
\newcommand{\Dn}{f(\Delta_{\mbox{\scriptsize n}})}
\newcommand{\Df}{f(\Delta_{\mbox{\scriptsize f}})}
\newcommand{\B}{\mbox{$B$}}
\newcommand{\Bo}{\mbox{$B$}_{0}}